# Thickness dependent tuning of the Berry curvature in a ferromagnetic Weyl semimetal

Yao Zhang[1,2,3†], Yuefeng Yin[4,5†], Guy Dubuis[1,2], Tane Butler[1], Nikhil V. Medhekar[4,5]* and Simon Granville[1,2]*

[1]Robinson Research Institute, Victoria University of Wellington, Wellington, New Zealand.
[2]MacDiarmid Institute for Advanced Materials and Nanotechnology, Wellington, New Zealand.
[3]School of Chemical and Physical Sciences, Victoria University of Wellington, Wellington, New Zealand.
[4]Department of Materials Science Engineering, Monash University, Clayton, Australia.
[5]ARC Centre of Excellence in Future Low Energy Electronics, Monash University, Clayton, Australia.

**Magnetic Weyl semimetals with spontaneously broken time-reversal symmetry exhibit a large intrinsic anomalous Hall effect originating from the Berry curvature. To employ this large Hall current for room temperature topo-spintronics applications, it is necessary to fabricate these materials as thin or ultrathin films. Here, we experimentally demonstrate that Weyl semimetal $Co_2MnGa$ thin films (20-50 nm) show a very large anomalous Hall angle ~11.4% at low temperature and ~9.7% at room temperature, which can be ascribed to the nontrivial topology of the band structure with large intrinsic Berry curvature. However, the anomalous Hall angle decreases significantly with thicknesses below 20 nm, which band structure calculations confirm is due to the reduction of the majority spin contribution to the Berry curvature. Our results suggest that $Co_2MnGa$ is an excellent material to realize room temperature topo-spintronics applications, however the significant thickness dependence of the Berry curvature has important implications for thin film device design.**

## Introduction

In Weyl semimetals (WSMs), a type of topological semimetal, the valence and conduction bands touch each other at isolated points called Weyl nodes, which can be understood as the monopoles and anti-monopoles of Berry curvature in momentum space [1]. Chiral Weyl fermions, as low-energy quasiparticle excitations around these Weyl nodal points, can be pumped along surface Fermi arcs that connect the paired Weyl nodes. Due to this unique nontrivial topology of band structure, WSMs display a rich variety of exotic transport properties [2,3], like negative magnetoresistance, and giant magnitudes of anomalous Hall effect (AHE), planar Hall effect, and anomalous Nernst effect (ANE).

Magnetic WSMs additionally have spontaneously broken time-reversal symmetry, resulting in large intrinsic AHE originating from the Berry curvature [4–6]. The intrinsic AHE current is dissipationless and fully spin-polarized and therefore has great potential for spintronics applications [7]. There is suggestion that magnetic WSMs could allow for the quantum anomalous Hall effect to persist at high temperatures [8]. Recently, kagome-lattice compound $Co_3Sn_2S_2$ has been proven to be a magnetic WSM with giant anomalous Hall conductivity $\sigma_{xy}$ ~1130 $\Omega^{-1}cm^{-1}$ and anomalous Hall angle, 20% [9], and thin films also exhibit the same transport behaviour as the bulk single crystal [10]. However, $Co_3Sn_2S_2$ shows a low Curie temperature $T_c$ = 177 K which means it cannot be used for room temperature devices. Co-based Heusler compounds, face-centered cubic metallic compounds with space-group symmetry $Fm\bar{3}m$(225), have been predicted to be promising candidates to realize magnetic WSMs with high Curie temperature [4,6]. Recent attention has also focused on magnetic Heusler alloys due to the wide tunability of the Berry curvature predicted in this materials class [2,11,12].

---

[†]These authors contributed equally to this work.

*Corresponding author. E-mail: Simon.Granville@vuw.ac.nz and nikhil.medhekar@monash.edu

What's more, these materials are soft ferromagnets, meaning that the spin orientation can be easily manipulated by applying a small magnetic field. Recently, bulk examples of full Heusler ferromagnet $Co_2MnGa$ with large Curie temperature $T_c \approx$ 700 K have been studied and shown to display giant ANE and large anomalous Hall angle, 12%, at room temperature, due to a large net Berry curvature near the Fermi energy associated with nodal lines and Weyl points [10,13,14]. In addition, Tung et al. [15] and Manna et al. [12], by calculating the Hall conductivity contribution of the majority and minority spin channels from bulk $Co_2MnGa$, found the spin-up and spin-down Hall currents would flow in opposite directions and the large $\sigma_{xy}$ results almost entirely from the majority spin channel. In this case, the anomalous Hall current would be nearly fully spin-polarized, even though $Co_2MnGa$ is not a half-metallic ferromagnet. The combination of topological electronic properties, ferromagnetism above room temperature, and strongly spin-polarized anomalous Hall current make $Co_2MnGa$ an exceptional candidate for studying the interplay of topology and magnetism and realizing room temperature topo-spintronics applications.

To realize these applications, $Co_2MnGa$ needs to be fabricated as thin films. There are studies of thin films prepared by molecular beam epitaxy [16–18], flash evaporation [19] or magnetron sputtering [20–22], however almost all of this work was done before the topological characteristics of this material were realised. Hence, there has so far been very little research about the transport properties related to the topological band structure in $Co_2MnGa$ thin films. Reichlova et al. found that $Co_2MnGa$ thin films exhibit a large ANE coefficient at 300 K, -2 to -3 $\mu$V/K, due to the nontrivial band structure[22,23]. However, this coefficient decreases significantly with thickness down to 10 nm and the reason still needs to be uncovered. Very recently, Markou et al. investigated thick $Co_2MnGa$ thin films which show the same value of anomalous Hall angle as in the single crystals [24], but they did not address the reason for the decline of anomalous Hall angle with decreasing thickness. Investigating the properties of ultrathin films is important for understanding the prospect of producing energy-efficient, high-density spintronic devices, as well as investigating the potential effects of their low dimensionality on the Berry curvature. In previous work we showed that ultrathin (>3.5 nm) $Co_2MnGa$ films display perpendicular magnetic anisotropy (PMA) in trilayer MgO/$Co_2MnGa$/Pd stacks [25], which could be used to reduce the switching current in spin-transfer torque devices [26].

For a comprehensive understanding of Weyl topological effects on the AHE in magnetic WSM thin films, we systematically investigated the transport properties of $Co_2MnGa$ thin films with thicknesses from 50 nm down to 5 nm. The thicker films exhibit a large anomalous Hall angle in agreement with the recent results of Markou et al., and ascribed to the intrinsic Berry curvature as found from calculations of bulk $Co_2MnGa$. However, for films thinner than 15 nm, the anomalous Hall angle drops significantly due to the reduction of the majority spin contribution to the Berry curvature, proving that tuning the Berry curvature can be achieved in this Heusler alloy and opening new possibilities for engineering topo-tronic materials and devices.

**Results**

**Crystal structure**

For the 50 nm film, the $\theta$-$2\theta$ scan of XRD shows the substrate MgO(002) peak and also Co2MnGa (002) and (004) peaks, as shown in Fig. 1(a), indicating a good out of plane texture. Other MgO[00$l$] lines are indicated by an asterisk, *. The inset is the rocking curve from the (004) peak, and the full width at half maximum (FWHM) is 0.779° suggesting a high crystal quality. Furthermore, the x-ray $\varphi$ scans from -45° to 315° [Fig. 1(b)] display a fourfold symmetry of the Co2MnGa (220) peak and the 45° rotation of these peaks relative to the substrate (220) peaks, confirming the epitaxial relationship of MgO(001)[100]||Co2MnGa(001)[110]. We also obtained the superstructure (111) peak with in-plane XRD which shows fourfold symmetry in the $\varphi$ scan as well, as shown in the top panel of Fig. 1(b). In full Heusler alloys, the XRD peaks with all odd (*hkl*) indexes, like (111), are known to originate in superlattice reflections in the $L2_1$ structure [27]. The in-plane $a$-$b$ and $c$-axis lattice constants are 5.765 Å and 5.759 Å ($c/a$ = 0.999), respectively, obtained from the (004) and (111) peaks. The strain $\varepsilon = (a_\| - a_0)/a_0$ is estimated as -0.035%, where $a_0$ is the lattice constant of bulk single crystal Co2MnGa (5.767 Å) [28] and $a_\|$ is the in-plane lattice constant of thin film (5.765 Å). Even though the diagonal $\sqrt{2}a_{\mathrm{MgO}}$ of the substrate is 5.962 Å which could induce a 3.27% lattice mismatch, the stress in our thin film is almost fully relaxed due to the post-annealing at 550 °C for 1 hour. The film roughnesses measured by XRR or AFM are below 1 nm, as shown by example for the 20 nm thick film in Fig. 1(c). Fitting the XRR results in a root-mean-square roughness (Rq) of 0.7 nm, with the thickness 20 nm ±2 nm. The AFM image confirms a smooth surface morphology with a Rq of 0.5 nm, as shown in the insert of Fig. 1(c).

**Magnetotransport properties**

Magnetic hysteresis loops of a 50 nm Co₂MnGa film were measured with the external magnetic field along both the in-plane and out-of-plane directions at 300 K, as shown in Fig. 2(a). These show the easy axis is along the Co₂MnGa[110] direction, the same as for other Heusler alloy thin films, Co₂MnAl Co₂MnSi [29,30], and the coercive field $\mu_0 H_C$ and saturation magnetization $M_s$ are 3 mT, 600 emu/cm$^{-3}$, respectively. Note that the crystalline direction used in this paper refers to the Co₂MnGa lattice, unless indicated otherwise. When the magnetic field is along the [100] direction, though it shows a small coercive field $\mu_0 H_c$ = 1 mT, the saturation field is 25 mT, larger than along the [110] direction. For the [001] direction, the saturation field is 1 T (see Figure S1), but it still shows a small square loop, interestingly indicating there is a small out-of-plane magnetization component at zero applied field. The uniaxial anisotropy energy density calculated from the hard [001] and easy [110] directions is $3.05 \times 10^6$ erg/cm³. The Hall resistivity as a function of magnetic field was measured at 300 K and 3 K, respectively, as shown in Fig. 2(b). For all of our samples, the current was applied along the [100] direction. The pure anomalous Hall resistivity was calculated by using $\rho_H^A = \rho_{xy} - R_o B$, where $R_o$ is the ordinary Hall coefficient and *B* is the magnetic flux density. The Hall coefficient $R_o$ is negative and quite small, -2.42×10⁻⁴ cm³/C (300 K) and -1.06×10⁻⁴ cm³/C (3 K), meaning that charge carriers are of the electron type. The anomalous Hall saturation fields at 300 K and 3 K are 1.3 T, 1.6 T respectively. The anomalous Hall resistivities are 17.1 $\mu\Omega\ cm$ and 15.4 $\mu\Omega\ cm$ at 300 K and 3 K, respectively, and the ordinary Hall resistivities at 9 T are 0.218 $\mu\Omega\ cm$ (300 K) and 0.095 $\mu\Omega\ cm$ (3 K). The Hall carrier concentration and Hall mobility are 2.58×10²² $cm^{-3}$ and 1.41 $cm^2 V^{-1} s^{-1}$, respectively, at 300 K. Since the ordinary Hall resistivity is negligible in our samples, it is safe to adopt $\rho_{xy}$ as nearly equal to the anomalous Hall resistivity. The knot in the resistivity curve also indicates the out-of-plane magnetization component seen in the magnetization measurements. In Fig. 2(c), the temperature dependence of Hall resistivity $\rho_{xy}$ and longitudinal resistivity $\rho_{xx}$ obtained from 300 K to 3 K are plotted. The 50 nm Co₂MnGa thin film has a residual resistivity $\rho_o$ = 136 $\mu\Omega\ cm$ at 23 K and exhibits a typical metallic behaviour at higher temperature (red curve). The $\rho_{xy}$ increases with *T*, displaying a maximum value 17 $\mu\Omega\ cm$ around 210 K, followed by a decrease to higher temperature (blue curve). The Curie temperature of Co₂MnGa is much higher than room temperature (see Figure S2), and consistent with the bulk $T_c$ ~ 700 K [31]. The Hall conductivity $\sigma_{xy}$ and longitudinal conductivity $\sigma_{xx}$ were calculated by using $\sigma_{xy} = \rho_{xy}/(\rho_{xy}^2 + \rho_{xx}^2(0\ T))$ and $\sigma_{xx} = \rho_{xx}/(\rho_{xy}^2 + \rho_{xx}^2(0\ T))$, respectively. Both Hall conductivity (blue curve) and longitudinal conductivity (red curve) increase monotonically on cooling and approach constant values of 812 $\Omega^{-1}cm^{-1}$ and 7250 $\Omega^{-1}cm^{-1}$, respectively, at low temperature, as shown in Fig. 2(d).

The anomalous Hall angle can be calculated by using $\theta_H = \sigma_{xy}/\sigma_{xx}$. Fig. 3(a) shows the anomalous Hall angle as a function of temperature from 300 K to 3 K. The 50 nm topological semimetal Co$_2$MnGa thin film shows a large anomalous Hall angle $\theta_H = 9.7\%$ at 300 K, increasing at low temperature to reach a maximum value 11.4% at 113 K. Usually, typical ferromagnetic thin films with topologically trivial band structure show a small angle, less than 3% [32]. Even though some non-magnetic Weyl semimetals exhibit giant anomalous Hall angle in single crystals [33,34], to date only Co$_2$MnGa shows such a large anomalous Hall angle in thin films at room temperature.

In non-magnetic WSMs, the existence of planar Hall effects or negative magnetoresistances has also been taken as further evidence of the effect of the Weyl character of the band structure on the electronic transport [35,36]. However, both of these effects are present in magnetic non-Weyl materials also, and the measurements of these effects in magnetic WSM Co$_3$Sn$_2$S$_2$ are unremarkable and consistent with a trivial magnetic origin [37]. In Figures 3 and 4 we show the angle-dependent planar Hall effect and magnetoresistance of our films measured along several crystallographic directions. The magnitudes of these effects are similar to those measured in Co$_3$Sn$_2$S$_2$ [9,37] and are not larger than in typical magnetic non-Weyl materials. What's more, the linear negative MR observed at high field in Co$_2$MnGa when the magnetic field is parallel to the current, as shown in Figure S4, can be ascribed to the thermal-induced spin disorder which is a typical ferromagnet behaviour [38,39]. Therefore, we consider the large anomalous Hall angle to be the only consistent magnetotransport indicator of the topological non-trivial band structure in ferromagnetic WSMs such as Co$_2$MnGa.

To understand the thickness dependence of the anomalous Hall angle, we fabricated Co$_2$MnGa films with smaller thicknesses, from 20 to 5 nm. Fig. 3(b) shows the thickness dependence of the anomalous Hall angle measured from 300 K to 3 K. Once we decrease the thickness to 20 nm, the anomalous Hall angle is still very large, around 11%, and this film shows a similar trend as the 50 nm film when decreasing the temperature. However, for samples with thicknesses of 15 nm, 10 nm and 5 nm, the anomalous Hall angle is temperature independent, and the value drops to 3.3%, 2.7% and 1.3%, respectively. The inset shows the maximum value of each sample, as measured between 3 K and 300 K. Our results of thinner samples agree very well with Ref. [39] in which the anomalous Hall angle of 6 nm Co$_2$MnGa is no more than ~2.8%. Fig. 3(c) is the log $\sigma_{xy}$ - $\sigma_{xx}$ plot for all samples measured at 300 to 3 K. $\sigma_{xx}$ is about 4.5 – 7.5×10$^3$ $\Omega^{-1}cm^{-1}$ at various thicknesses and temperatures, and $\sigma_{xy}$ is nearly independent of $\sigma_{xx}$. For the samples with thicknesses from 50 nm to 20 nm, one can see that $\sigma_{xy}$ is nearly constant (around 700 $\Omega^{-1}cm^{-1}$). It is known that there exist three regimes in the scaling behaviour of the AHE depending on the range of $\sigma_{xx}$, categorized by a scaling parameter, $\alpha$, in $\sigma_{xy} \propto \sigma_{xx}^\alpha$ [40]. The values of $\sigma_{xx}$ of all samples are close to the intermediate region as defined in Ref. [40], where $\sigma_{xx} = 10^4 – 10^6$ $\Omega^{-1}cm^{-1}$. Furthermore, this constant value of $\sigma_{xy}$ is close to the order of $e^2/ha$ ~ 670 $\Omega^{-1}cm^{-1}$, where $h$ is Planck's constant and $a$ is lattice constant. This is the value expected from the intrinsic scattering-independent mechanism originating from the Berry curvature in momentum space, and indicates that $\sigma_{xy}$ in our films is dominated by the intrinsic component due to the nontrivial topology of band structure of Co$_2$MnGa [32]. However, for the thinner samples with thicknesses of 15 nm, 10 nm and 5 nm, the values of $\sigma_{xy}$ drop steeply to 170, 160 and 70 $\Omega^{-1}cm^{-1}$, respectively. Since the values of $\sigma_{xy}$ of these samples are much smaller than $e^2/ha$, the $\sigma_{xy}$ might from not be due to only the intrinsic scattering, but also might include contributions from the extrinsic side-jump contributions which also can induce value of $\sigma_{xy}$ that is independent of $\sigma_{xx}$ [41]. This agrees with our previous results that the $\sigma_{xy}$ of ultrathin Co$_2$MnGa films in trilayer stacks results from both intrinsic and side-jump scattering [25]. Fig. 3(d) and (e) show the $\sigma_{xx}$ and $\sigma_{xy}$ against the temperature and thickness, respectively.

The $\sigma_{xx}$ slightly increases with decreasing temperature, and then remains constant at low temperature. $\sigma_{xx}$ for all samples at various temperatures varies only by approximately 25%, however, the $\sigma_{xy}$ drops suddenly once the thickness is below 15 nm. For instance, comparing the 50 nm and 10 nm films, the $\sigma_{xx}$ of both is around $7.2\times10^3$ $\Omega^{-1}cm^{-1}$ at 3 K, but the $\sigma_{xy}$ drops from 810 to 160 $\Omega^{-1}cm^{-1}$ for the thinner film.

**Evolution of Berry curvature and surface band structure**

The intrinsic contribution to the $\sigma_{xy}$ is dependent only on the band structure of a material. It can be calculated from the Kubo formula for the Hall conductivity

$$\sigma_{xy} = \frac{e^2}{\hbar}\int \frac{d\mathbf{k}}{2\pi^3}f(\varepsilon_k)\mathbf{\Omega}_z,$$

where $f(\varepsilon_k)$ is the Fermi-Dirac distribution function, $\mathbf{\Omega}_z$ is the z-component Berry curvature [42]. The $\sigma_{xy}$ can also be found from the sum of the conductivities for majority spin electrons (↑) and minority spin electrons (↓), $\sigma_{xy} = \sigma_{xy}^{\uparrow} + \sigma_{xy}^{\downarrow}$. Thus, the spin polarization-dependent conductivities $\sigma_{xy}^{\uparrow}$ and $\sigma_{xy}^{\downarrow}$ can be calculated by taking the integral of $\mathbf{\Omega}_z^{\uparrow}$ and $\mathbf{\Omega}_z^{\downarrow}$ over the entire Brillouin zone (BZ). Manna *et al.*, by calculating the Hall conductivity contribution of the majority and minority spin channels from bulk Co$_2$MnGa, found $\sigma_{xy}$ ~ 1600 $\Omega^{-1}cm^{-1}$ and $\sigma_{xy}^{\downarrow}$ is almost equal to 0 at the Fermi level ($E_F$) [12]. The value of $\sigma_{xy}$ in our thin films is half that of bulk Co$_2$MnGa. However, interestingly we have almost the same anomalous Hall angle of the bulk reports, 12%, since the value of $\sigma_{xx}$ in our thin film is relatively smaller than in Ref. [12].

To understand the influence of spin-dependent electrons for $\sigma_{xy}$ of Co$_2$MnGa thin films with different thickness, we calculate the *z*-component of the Berry curvature from majority spin electrons and minority spin electrons near the Fermi level in the $k_x$-$k_y$ plane, respectively, as shown in Fig. 4. For 123 primitive cell (50.1 nm) thick slab of Co$_2$MnGa, $\mathbf{\Omega}_z^{\uparrow} \gg \mathbf{\Omega}_z^{\downarrow}$ (see Fig. 4(b)(d)), meaning $\sigma_{xy}^{\uparrow} \gg \sigma_{xy}^{\downarrow}$ and suggesting that the Berry curvature contribution to $\sigma_{xy}$ from the majority spin electrons is much higher than that from minority spin electrons, which is the same as for bulk Co$_2$MnGa. The magnitude of $\mathbf{\Omega}_z^{\uparrow}$ of this thick Co$_2$MnGa film is close to bulk values and is consistent with the $\sigma_{xy}$ values calculated by others[10,12,15], suggesting films of ~50 nm thickness still show the bulk-like properties, like the large $\sigma_{xy}$ originating from the nontrivial topology of the band structure with large intrinsic Berry curvature. The $\mathbf{\Omega}_z^{\uparrow}$ for the 50 nm film in Fig. 4 is complicated because it includes multiple 'slabs' in the calculation, and at 50 nm there is not yet enough overlap of the states so that continuous bulk-like bands are formed. Once we decrease the thickness to 4.9 nm (12 primitive cells), both $\mathbf{\Omega}_z^{\uparrow}$ and $\mathbf{\Omega}_z^{\downarrow}$ drop significantly (see Fig. 4(a)-(d)). This agrees with the experimental observation that the $\sigma_{xy}$ reduces when film thickness decreases. Figs. 4(e) and (f) summarise the $\mathbf{\Omega}_z^{\uparrow}$ and $\mathbf{\Omega}_z^{\downarrow}$ as a function of *k* along the Γ-M line for various thicknesses. The extent of Berry curvature reduction for majority spins is more significant than that for minority spins. Once the thickness is down to 20 nm, the $\mathbf{\Omega}_z^{\uparrow}$ slightly drops, but it is still close to the bulk value indicating that the $\sigma_{xy}$ is robust down to this thickness. Interestingly, the Berry curvature contribution from minority spin electrons is dominating in some regions for 4.9 nm Co$_2$MnGa which is an important qualitative change from the bulk one. Note the very different maxima used in the scales of Fig. 4(a)-(d) vs Fig. 4(e)-(f), so that the main features can be seen in Fig. 4(a)-(d), and the very large spikes can be seen in Fig. 4(e)-(f).

**Discussion**

By comparing to previously reported results of thin films, including Heusler alloys and classic ferromagnetic transition metal and alloys, the anomalous Hall angle in Co$_2$MnGa shows a large value, 11.4%, and also a comparatively large Hall conductivity 775 $\Omega^{-1}cm^{-1}$, as shown in Fig. 5. To obtain a large anomalous Hall angle, the $\sigma_{xy}$ needs to be large, or $\sigma_{xx}$ needs to be small, but large $\sigma_{xy}$ signal with very good signal-to-noise ratio is necessary for logic or memory hybrid CMOS/AHE devices [43] and AHE sensors [44]. Near the dirty regime [42] with a small $\sigma_{xx}$, however, $\sigma_{xy}$ is suppressed more drastically than $\sigma_{xx}$, due to quasiparticle damping or spoiling of the resonance condition [45]. For ferromagnetic systems, both $\sigma_{xy}$ and $\sigma_{xx}$ are either large or small, and therefore the anomalous Hall angle is usually not very large. Though Co$_3$Sn$_2$S$_2$ thin films have a giant anomalous Hall angle 20%, it shows a low Curie temperature $T_c$ = 177 K as we mentioned before. Even for exceptions such as the compensated Heusler ferrimagnet Mn$_{2-x}$Ru$_x$Ga, which has a large anomalous Hall angle 7.7% at 300 K, it displays a small $\sigma_{xy}$, around 220 $\Omega^{-1}cm^{-1}$ [46].

In summary, we have prepared high quality epitaxial films of the topological semimetal Co$_2$MnGa with thicknesses from 5 nm to 50 nm on MgO substrates and established they have a similar large anomalous Hall angle at room temperature as in bulk crystals. For the 50 nm thick film, it shows large $\sigma_{xy}$ ~ 777 $\Omega^{-1}cm^{-1}$ and anomalous Hall angle 11.4% at 113 K. The origin of the large $\sigma_{xy}$ is the large net Berry curvature near the $E_F$ due to the nontrivial topology of band structure of Co$_2$MnGa, and still shows a large anomalous Hall angle 9.7% at 300 K. The bulk-like properties are robust, for film thicknesses down to 20 nm. Once we decrease thickness below 20 nm, the $\sigma_{xy}$ suddenly drops a lot with a slight change of $\sigma_{xx}$, with a corresponding decrease in the anomalous Hall angle. From our band structure calculations, we ascribe the decrease to the reduction of the majority spin electron Berry curvature. Our results confirm that Co$_2$MnGa is a topological magnetic material in which tuning of the Berry curvature is possible, as observed in the anomalous Hall behaviour. The opportunity to control the Berry curvature in materials provides a way to realize topo-tronic logic or memory devices and AHE sensors that operate at room temperature.

## Methods

**Sample preparation**

Co$_2$MnGa (5-50 nm) thin films were epitaxially deposited on single-crystalline MgO(001) substrates in a Kurt J Lesker CMS-18 magnetron sputtering system with a base pressure below 5×10$^{-8}$ Torr. Before fabricating thin films, substrates were cleaned with an Ar plasma, and then annealed at 400 $^o$C for 1 hour in the vacuum chamber. Co$_2$MnGa thin films were DC magnetron sputtered from a stoichiometric polycrystalline target at 100 W under 6 mTorr of Ar with a growth rate of 0.8 Å/s at 400 $^o$C while rotating the sample holder. All samples were post-annealed *in situ* at 550 $^o$C for 1 hour to improve the lattice structure and the atomic ordering among Co, Mn, and Ga sites. After cooling down to ambient temperature, a 2 nm protective MgO layer was deposited on the top. Samples were patterned into standard Hall bars (*l* = 2000 $\mu$m, *w* = 150 $\mu$m) by photolithography and then Ar ion milling for magnetotransport characterization.

**Thin film characterization**

The crystalline structure of the Co$_2$MnGa thin films was characterized by x-ray diffraction (XRD) with Co $K_\alpha$ radiation ($\lambda$ = 1.7889 Å) using a Bruker D8 Advance. The film thicknesses were checked by x-ray reflectivity (XRR) using a PANalytical X'Pert PRO. Atomic force microscopy (AFM) was carried out on a Nanosurf FlexAFM mounted on a Nanosurf Isostage active vibration cancellation stand.

**Magnetization and electric transport measurements**

Magnetization measurements were performed using the RSO module of a Superconducting Quantum Interference Device magnetometer (SQUID, Quantum Design). The transport measurements were performed using the resistivity option of a Physical Property Measurement System (PPMS, Quantum Design).

**Band structure calculation method**

We used density functional theory (DFT) calculations as implemented in the Vienna ab initio Simulation Package (VASP) to calculate the optimized geometry and the electronic structure of $Co_2MnGa$ [47]. The Perdew-Burke-Ernzehof (PBE) form of the generalized gradient approximation (GGA) was used to describe electron exchange and correlation [48]. The kinetic energy cutoff for the plane-wave basis set was set to 400 eV. We used a 12 × 12 × 12 Γ-centered $k$-point mesh for sampling the Brillouin zone. For calculation of Berry curvature of $Co_2MnGa$ thin films, we adopted the method as implemented in the open-source code WannierTools [49,50], based on the Wannier tight-binding Hamiltonian obtained from wannier90 [51]. Co/Mn $d$ and Ga $s$, $p$ orbitals are projectors for tight-binding Hamiltonian construction.

**Data availability**

The data that support the findings of this study are available from the corresponding authors upon reasonable request.

**Acknowledgments**

We are grateful to Jérôme Leveneur from GNS science for the AFM measurements and Sarah Spencer from the Robinson Research Institute for EDX. Y.F.Y and N.M acknowledge the support from Australian National Computing Infrastructure and Pawsey Supercomputing Centre. The MacDiarmid Institute is supported under the New Zealand Centres of Research Excellence Programme. Y.F.Y and N.M are thankful for the funding support from the Australian Research Council (CE170100039).

**Author contributions**

Y.Z fabricated and characterized devices with assistance from G.D and T.B. Y.Z and S.G analyzed the data. Y.F.Y did the theoretical calculations with contributions from N.M. Y.Z and S.G. wrote the manuscript with contributions from all authors. All authors discussed the results and commented on the manuscript. The study was performed under the supervision of S.G.

**Figures**

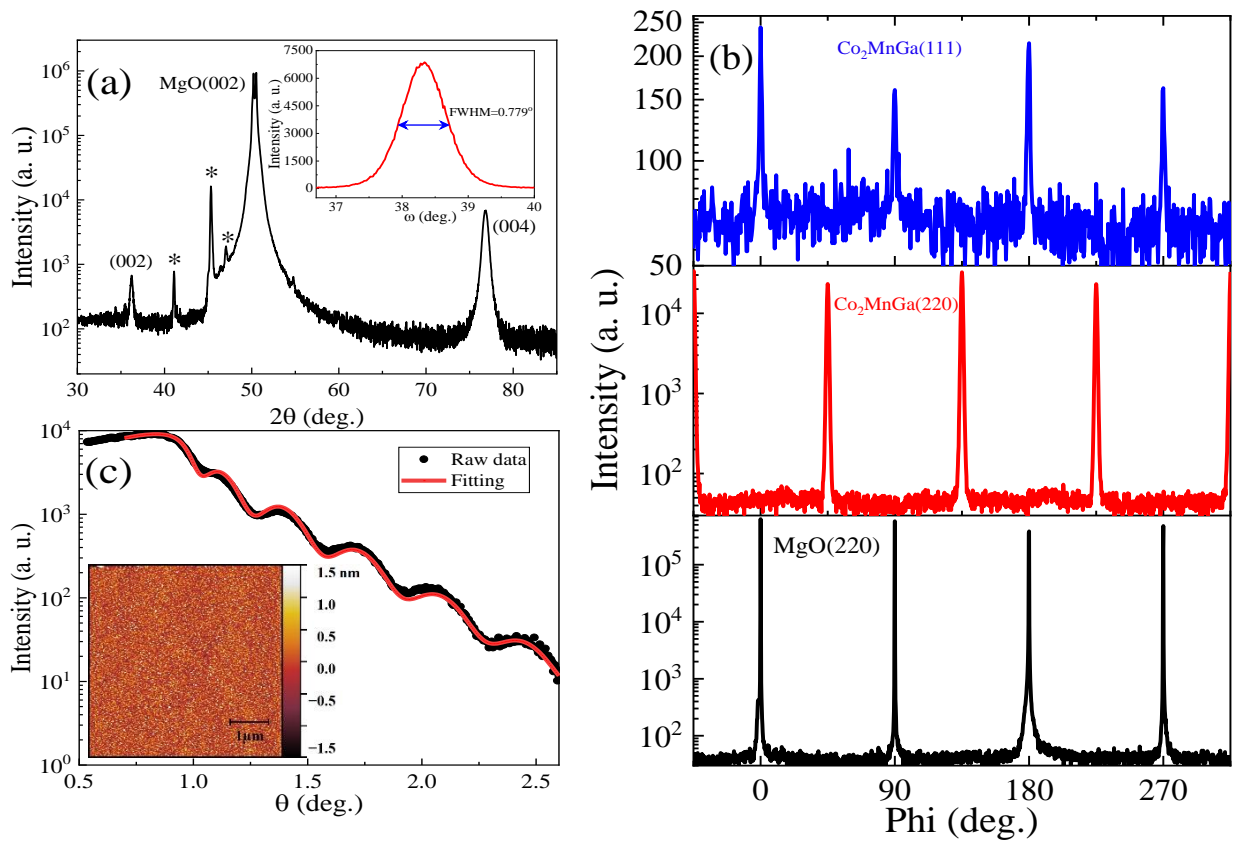

**Fig. 1. Crystal structure and surface morphology of Co₂MnGa.** (a) $\theta$-$2\theta$ scan, the inset is the rocking curve from (004) peak, and (b) in plane $\varphi$ scans from -45° to 315°, for 50 nm Co₂MnGa thin films. (c) X-ray reflectivity for the 20 nm Co₂MnGa thin film, the inset is the atomic force microscopy image (5 × 5 $\mu$m²) with colour contrast (dark to bright) for the height scale corresponds to 3 nm.

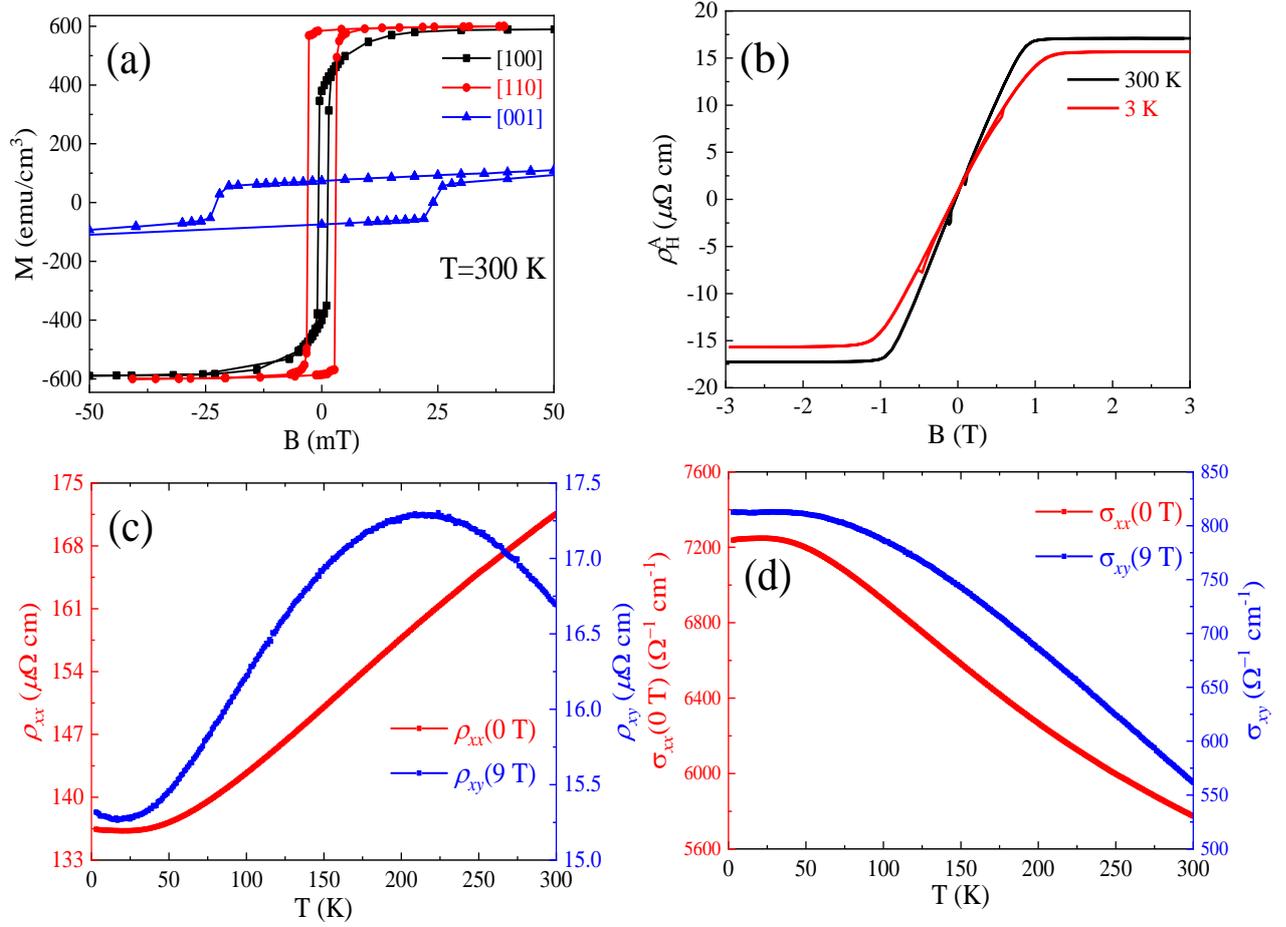

**Fig. 2. Magnetic and electric transport measurements for 50 nm Co₂MnGa.** (a) Magnetization measured at 300 K with the magnetic field along in-plane and out-of-plane direction, respectively. (b) Anomalous Hall effect measurements at 3 K and 300 K, respectively. (c) Longitudinal resistivity without field (red) and Hall resistivity cooling down with 9 T (blue) as a function of temperature from 3 K to 300 K. (d) the corresponding conductivity as a function of temperature from 3 K to 300 K.

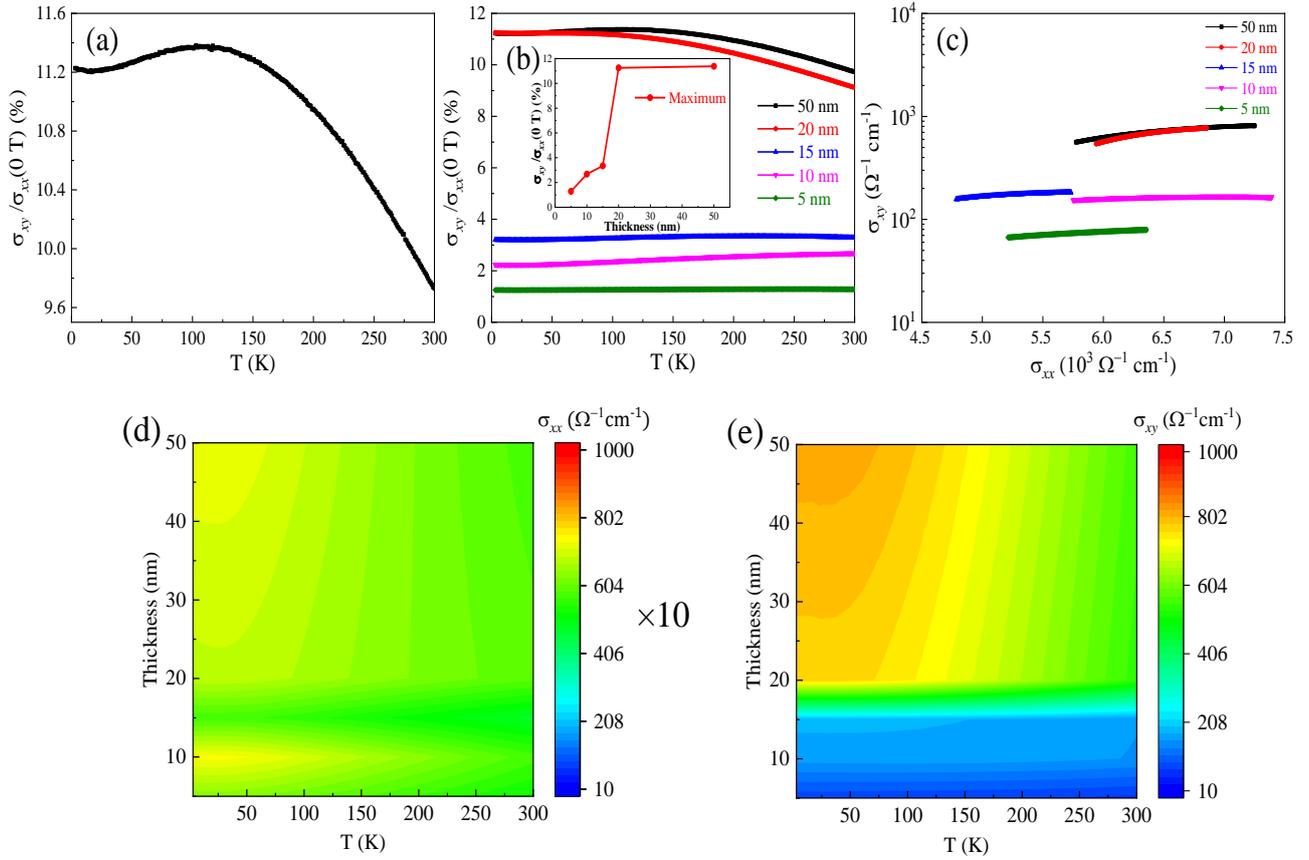

**Fig. 3. Transport measurement of the anomalous Hall angle.** (a) Temperature dependent anomalous Hall angle measured from 3 K to 300 K for 50 nm $Co_2MnGa$. (b) Thickness dependence of anomalous Hall angle as a function of temperature from 3 K to 300 K, the inset is the maximum anomalous Hall angle as a function of thickness. (c) Log $\sigma_{xy}$ - $\sigma_{xx}$ plot for $Co_2MnGa$ with thicknesses from 5 nm to 50 nm. (d)(e) Thickness dependence of $\sigma_{xx}$ and $\sigma_{xy}$ as a function of temperature from 3 K to 300 K, respectively.

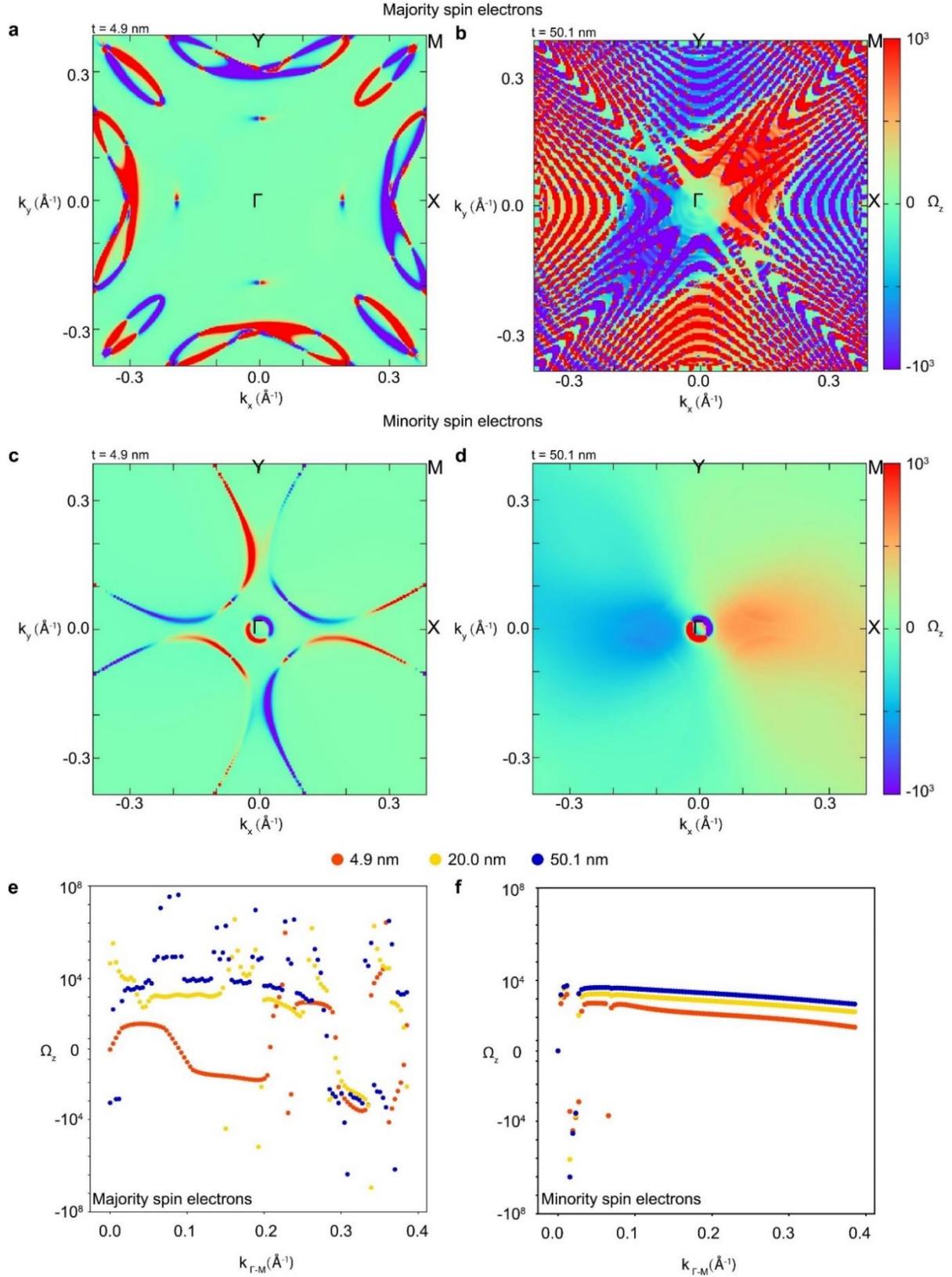

**Fig. 4. Evolution of Berry curvature in Co$_2$MnGa thin films.** (a)-(d) show the calculated integral of z-component Berry curvature ($\mathbf{\Omega}_z$) of occupied bands from minority and majority spin electrons for Co$_2$MnGa thin films of 4.9 nm and 50.1 nm, (e) and (f) show a comparison of $\mathbf{\Omega}_z$ along Γ-M line for Co$_2$MnGa thin films of 4.9 nm, 20.0 nm and 50.1 nm.

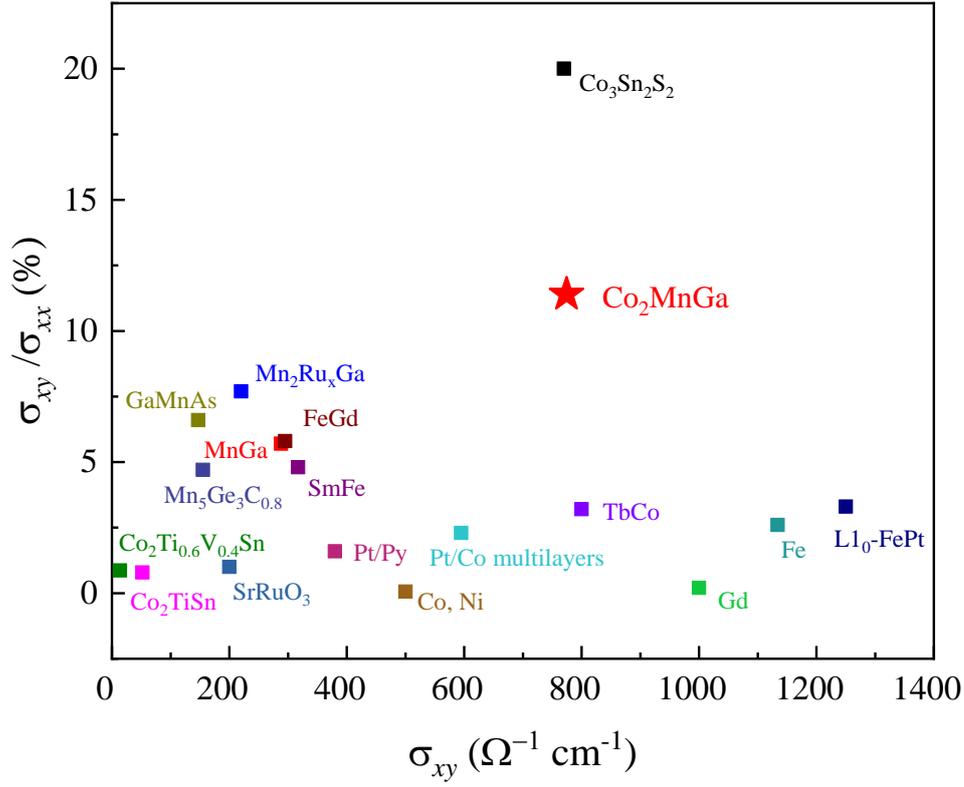

Fig. 5. The $\sigma_{xy}$ dependent anomalous Hall angle of our result in comparison with previously reported results for other thin films.